\documentclass[twocolumn,aps,prl,preprintnumbers,amsmath,amssymb,showpacs]{revtex4-1}
\usepackage{bbm}
\usepackage{mathrsfs}
\usepackage{amssymb,amsmath}
\usepackage[dvips]{graphicx}
\usepackage{epsfig}
\usepackage{color}
\usepackage{ulem}

\usepackage[utf8]{inputenc}

\newcommand{\textout}[1]{}

\newcommand{\veps}{\varepsilon}

\newcommand{\munichBloch}{Physics Department, Quantum Optics Chair, Ludwig-Maximilians-Universit\"at M\"unchen, 80799 M\"unchen, Germany}
\newcommand{\berlin}{Institute of Advanced Study Berlin, 14193 Berlin, Germany}
\newcommand{\garchingBloch}{Max-Planck Institut f\"ur Quantenoptik, 85748 Garching, Germany}
\newcommand{\munich}{Physics Department, Arnold Sommerfeld Center for Theoretical Physics and Center for NanoScience, Ludwig-Maximilians-Universit\"at M\"unchen, 80333 M\"unchen, Germany}

\newcommand{\queensland}{School of Physical Sciences, The University of Queensland, Brisbane, QLD 4072, Australia}
\newcommand{\paris}{Laboratoire Mat\' eriaux et Ph\' enom\`enes Quantiques,
Universit\' e Paris Diderot-Paris 7 and CNRS, UMR 7162, 75205 Paris Cedex 13, France}

\begin{document}
\title{Landau-Zener sweeps and sudden quenches in coupled Bose-Hubbard chains}
\author{C. Kasztelan$^{1}$, S. Trotzky$^{2,3}$, Y.-A. Chen$^{2,3}$, I. Bloch$^{2,3}$, I. P. McCulloch$^{4}$, U. Schollw\"ock$^{1,5}$ and G. Orso$^{1,6}$} 
\affiliation{$^{1}$ \munich}
\affiliation{$^{2}$ \munichBloch}
\affiliation{$^{3}$ \garchingBloch}
\affiliation{$^{4}$ \queensland}
\affiliation{$^{5}$ \berlin}
\affiliation{$^{6}$ \paris}

\begin{abstract}
We simulate numerically the dynamics of strongly correlated bosons in a two-leg ladder subject to a time-dependent energy bias between the two chains. When all atoms are initially in the leg with higher energy, we find a drastic reduction of the inter-chain particle transfer for slow linear sweeps, in quantitative agreement with recent experiments. This effect is preceded by a rapid broadening of the quasi-momentum distribution of atoms, signaling the presence of a bath of low-energy excitations in the chains. We further investigate the scenario of quantum quenches to fixed values of the energy bias. We find that for large enough density the momentum distribution relaxes to that of an equilibrium thermal state with the same energy.
\end{abstract}


\date{\today}

\maketitle

The investigation of dynamical phenomena in quantum many-body systems provides an efficient way to probe the physical properties of the underlying ensemble and to tackle fundamental questions of statistical mechanics. Of special interest in this regard is how the quantum dynamics of two coupled modes is changed in a many-body setting \cite{Zobay:2000, LiuFu:2002, Witthaut:2006, Venumadhav:2010} and how a closed quantum system far from equilibrium does or does not equilibrate (see Refs.~\cite{Rigol:2008, Barthel:2009, Cramer:2010, Polkovnikov:2010} and references therein). The high degree of control achieved in current experiments with ultracold atoms in optical lattices \cite{Bloch:2008} allows for clean and quantitative studies of quantum many-body dynamics \cite{Greiner:2002a, Fertig:2005, Kinoshita:2006, Tuchman:2006, Sebby:2007a, Will:2010, Trotzky:2011}. Recently, a generalization of the famous Landau-Zener (LZ) sweep \cite{Landau:1932,*Zener:1932} to a pair of coupled one-dimensional quantum gases of strongly correlated bosonic particles was addressed experimentally \cite{Chen:2010}. One intriguing result of this experiment was the observation of a breakdown of adiabatic inter-chain transfer in slow sweeps of the inter-chain bias.

In this Letter, we present a numerically exact study of the LZ dynamics in two coupled Bose-Hubbard chains subject to a time-dependent inter-chain bias energy, using the time-dependent density matrix renormalization group ($t$-DMRG) method \cite{White:2004, *Daley:2004, *Schollwoeck:2011}. We show that the breakdown of adiabatic transfer as observed in Ref.~\cite{Chen:2010} in slow sweeps far away from the ground-state is always accompanied by a dramatic broadening of the quasi-momentum distribution of atoms in the two legs. This provides strong evidence that the responsible mechanism for the breakdown is the coupling to an internal bath of low-energy momentum excitations. Finally, we study the quantum dynamics emerging after sudden quenches of the bias energy where the same mechanism causes a relaxation to a steady state similar to an equilibrium thermal state with the same energy.

\textit{Setup and model}.
We consider a two-leg ladder formed by two coupled Bose-Hubbard chains [see Fig.~\ref{fig:inverseSweep}(a)] described by the Hamiltonian 
\begin{eqnarray}\label{Ham}
\hat H(t)&=&-\bigg(J\!\sum_i  \hat b^\dag_{i,{\rm L}}  \hat b_{i,{\rm R}} +J_{\|}\sum_{i,\sigma}\!  \hat b^\dag_{i,\sigma} \hat b_{i+1,\sigma}\bigg) +h.c. \nonumber \\
&&+\frac{U}{2}\sum_{i,\sigma} \hat n_{i,\sigma}( \hat n_{i,\sigma}-1) 
  +\Delta(t)\sum_{i}  \hat n_{i,{\rm R}},
\end{eqnarray}
where $\hat b_{i,\sigma} $ ($i=1\ldots L_s$, $\sigma = {\rm L},{\rm R}$) are the local annihilation operators and  $\hat n_{i,\sigma}=\hat b^\dag_{i,\sigma} \hat b_{i,\sigma}$. The parameters $J_{\|}$ and $J$ are the intra-chain and inter-chain tunnel couplings, respectively, $U$ is the on-site interaction energy, and $\Delta(t)$ is a time-dependent energy bias between the two chains. In the following we fix $J=\hbar=1$ and we call $n = N/L_s$ the density of the system with $N=\sum_{i,\sigma} \langle \hat n_{i,\sigma}\rangle$. The zero-temperature properties of the model (\ref{Ham}) have been investigated recently using static DMRG and mean-field methods \cite{Danshita:2007}.

\begin{figure}[tb]
\centering
\includegraphics[scale=1]{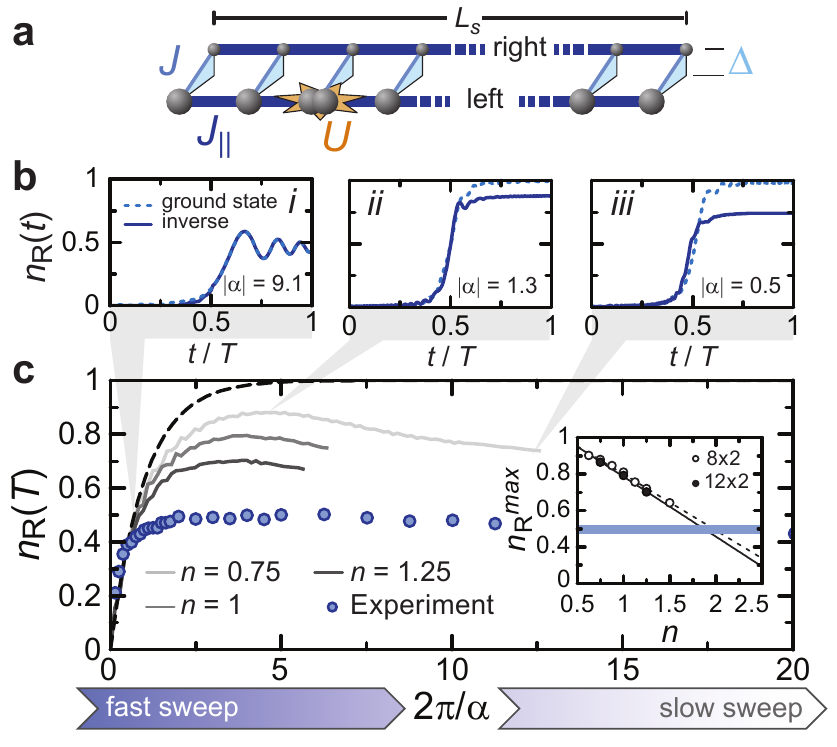}
\caption{\label{fig:inverseSweep}
(Color online) (a) Sketch of the two-leg ladder described by the Hamiltonian Eq.~(1) (b) Instantaneous right-leg population $n_{\rm R}(t)$ during ground-state (dashed lines) and inverse sweeps (solid lines) with various rates $\alpha$, calculated with $n = 0.75$ and $L_s=8$. The Hamiltonian parameters are $|\Delta_0|=18.2$, $J_\parallel=0.38$, and $U=1.58$. (c) Transfer efficiency $n_{\rm R}(T,2\pi/\alpha)$ for inverse sweeps with the same parameters and $n=0.75, 1, 1.25$ (solid lines). The dashed line is the two-level LZ probability $p_{\rm LZ}(\alpha)$. Open circles represent experimental data taken from Ref.~\cite{Chen:2010} for the corresponding set of control parameters [see Appendix B]. The inset shows the maximum transfer efficiency as a function of the density for $L_s=8$ (open circles) and $12$ (filled circles) and the linear extrapolation to higher densities (dashed and solid line, respectively). The blue line marks the maximum reached in the experiment.}
\end{figure}

We initialize the system in the ground state of Hamiltonian Eq.~(\ref{Ham}) with a large positive bias $\Delta(t=0) \rightarrow +\infty$ so that all particles are initially located in the left leg. This ground state is found from a static DMRG calculation. We distinguish two different scenarios for the time-dependence of the bias $\Delta(t)$:
\begin{equation}
  \Delta(t>0) =
\begin{cases}
\Delta_0 + \alpha\,t\,,\;\;\; & \mbox{(linear sweep)}\label{deltat} \\
\Delta_f, &\mbox{(sudden quench).}  
\end{cases}
\end{equation} 
The first case corresponds to a linear LZ sweep $\Delta_0 \rightarrow -\Delta_0$ of the bias with constant rate $\alpha = -2\Delta_0/T$,  $T$ being the sweep time. The second case amounts to a sudden change (quench) of the bias to a constant value $\Delta_f$. We use $t$-DMRG \cite{White:2004,*Daley:2004,*Schollwoeck:2011} to calculate the exact quantum dynamics in both situations for system sizes up to $L_s = 16$. In all the simulations we fix the truncation error to $\veps= 10^{-4}$ and check that all results have converged. 

\textit{Linear sweeps.} We first discuss LZ sweeps of the energy bias, focusing on \textit{inverse sweeps}, where the left leg is initially higher in potential energy than the right leg ($\Delta_0 < 0$, $\alpha > 0$). The corresponding counterpart we call a \textit{ground-state sweep} ($\Delta_0 > 0$, $\alpha < 0$). In the limit\textout{where} $L_s = 1$ and $U=J_{\|}=0$, the problem reduces to the original two-level LZ problem \cite{Landau:1932, *Zener:1932}, where the probability to reach the right-leg as a function of the sweep rate $p_{\rm LZ}(\alpha) = 1-\exp(-2\pi/\alpha)$ is the same for both.

We consider the fraction $n_{\rm R}(t)=N^{-1}\sum_i \langle \hat  n_{i,{\rm R}} (t)\rangle$ of particles in the right leg at a given time $t$. This is plotted in Fig.~\ref{fig:inverseSweep}(b) for both types of sweeps and three absolute values of the sweep rate $\alpha$. In Fig.~\ref{fig:inverseSweep}(c), we plot the final transfer efficiency $n_{\rm R}(T)$ reached at the end of the inverse sweep as obtained from a number of such traces with varying sweep rates [see Appendix A]. We compare the results for $n=0.75, 1, 1.25$ to experimental data taken from Ref.~\cite{Chen:2010} for the respective set of Hamiltonian parameters [see Appendix B]. In the experiments, the effective density was significantly higher as will be discussed in detail below.

For fast sweeps, corresponding to $|\alpha|/2 \pi  \gg U, J_{\|}$, the curves for the ground-state and for the inverse sweep lie on top of each other [Fig.~\ref{fig:inverseSweep}(b)-\textit{i}]. This is due to the fact that practically no intra-chain processes can occur during the sweep and hence the problem is effectively reduced to the original two-level LZ problem. As a consequence the DMRG results in Fig.~\ref{fig:inverseSweep}(c) collapse onto $p_{\rm LZ}(\alpha)$ for small values of $2\pi/\alpha$.

Differences between the two kinds of sweeps become evident as $\alpha$ decreases and intra-chain processes start to affect the dynamics. Here, we find the transfer efficiency for the inverse sweeps to be reduced relative to the ground-state sweep [Fig.~\ref{fig:inverseSweep}(b)-\textit{ii}]. This can in parts be understood in the limit of isolated double wells filled with interacting particles ($J_\parallel =0$, $U>0$). Here, the dynamical problem can easily be solved numerically by direct integration of the Schr\"odinger equation and even analytically when assuming $U \gg 1$ \cite{Venumadhav:2010, Chen:2010}. The analysis of this scenario shows that with increasing number of particles the transfer efficiency is enhanced in the ground-state sweep, while it is reduced in the inverse sweep [see Appendix C].

For slow enough sweeps, corresponding to $\vert\alpha\vert/2\pi \lesssim \textrm{min}(U,J_{\|})$, the transfer efficiency for the inverse sweep starts to decrease as the rate $\alpha$ decreases, as it was found in Ref.~\cite{Chen:2010}. This is in sharp contrast with the behavior for the ground-state sweep, where the transfer efficiency approaches unity [Fig.~\ref{fig:inverseSweep}(b)-\textit{iii}]. We emphasize that the breakdown of adiabaticity in the inverse sweep cannot be obtained within the isolated double-well picture and is directly related to the possibility for the system to generate excitations along the chains. Since these excitations have to be created via collisions, the breakdown is more effective in systems with higher densities, as it is evident in Fig.~\ref{fig:inverseSweep}(c). Eventually, for infinitely slow sweeps ($\alpha \to 0$) we expect the transfer efficiency to rise to 1 again due to the finiteness of the system.

We note, that the experimental results taken for comparison were obtained from a two-dimensional array of pairwise coupled chains with \textit{inhomogeneous} density \cite{Chen:2010}. In each of these ladders the maximum fidelity is reached for different sweep rates and takes different values. Therefore, the maximum in $n_{\rm R}(T)$ is less pronounced than in the numerical results for a single ladder. From ground-state DMRG calculations respecting the experimental geometry, we find that the average density in the center of the chains was $\langle n(z=0)\rangle_\textrm{av} \simeq 2$, whereas the overall average density was $\langle n\rangle_\textrm{av} \simeq 1.4$. A rapid growth in entanglement entropy \cite{Calabrese:2005,Osborne:2006} prevents us to access the dynamics with these densities -- or of the full inhomogeneous system -- directly in the simulations. In the inset of Fig.~\ref{fig:inverseSweep}(c) we plot the maximum efficiency $n_{{\rm R}, {\rm max}}$ reached in the DMRG simulations as a function of $n$ and for $L_s=8, 12$. We find $n_{{\rm R}, {\rm max}}$ to depend approximately linearly on $n$ and only weakly on $L_s$.  Linear extrapolation of our numerical results for $n_{{\rm R},{\rm max}}$ to higher densities yields a crossing with the experimentally measured value at $n \simeq 1.8$.

Having recovered the downturn of the transfer efficiency observed in the experiments, we now turn to the discussion of the experimentally accessible quasi-momentum distribution of atoms in both legs \cite{Bloch:2008}. The latter is defined as $n_{k\sigma} = L^{-1} \sum_{m,s}e^{-ik(m-s)}\langle b_{m\sigma}^\dag b_{s\sigma}\rangle$, where the sum extends to all lattice sites and $k$ is given in units of the reciprocal lattice constant. In Fig.~\ref{fig:MomentumDistrib}(a), we show $n_{k\sigma}$ for the left and the right leg as a function of $2\pi/\alpha$ and $k$ and their respective widths in Fig.~\ref{fig:MomentumDistrib}(b).

\begin{figure}[tb]
\centering
\includegraphics[scale=0.55]{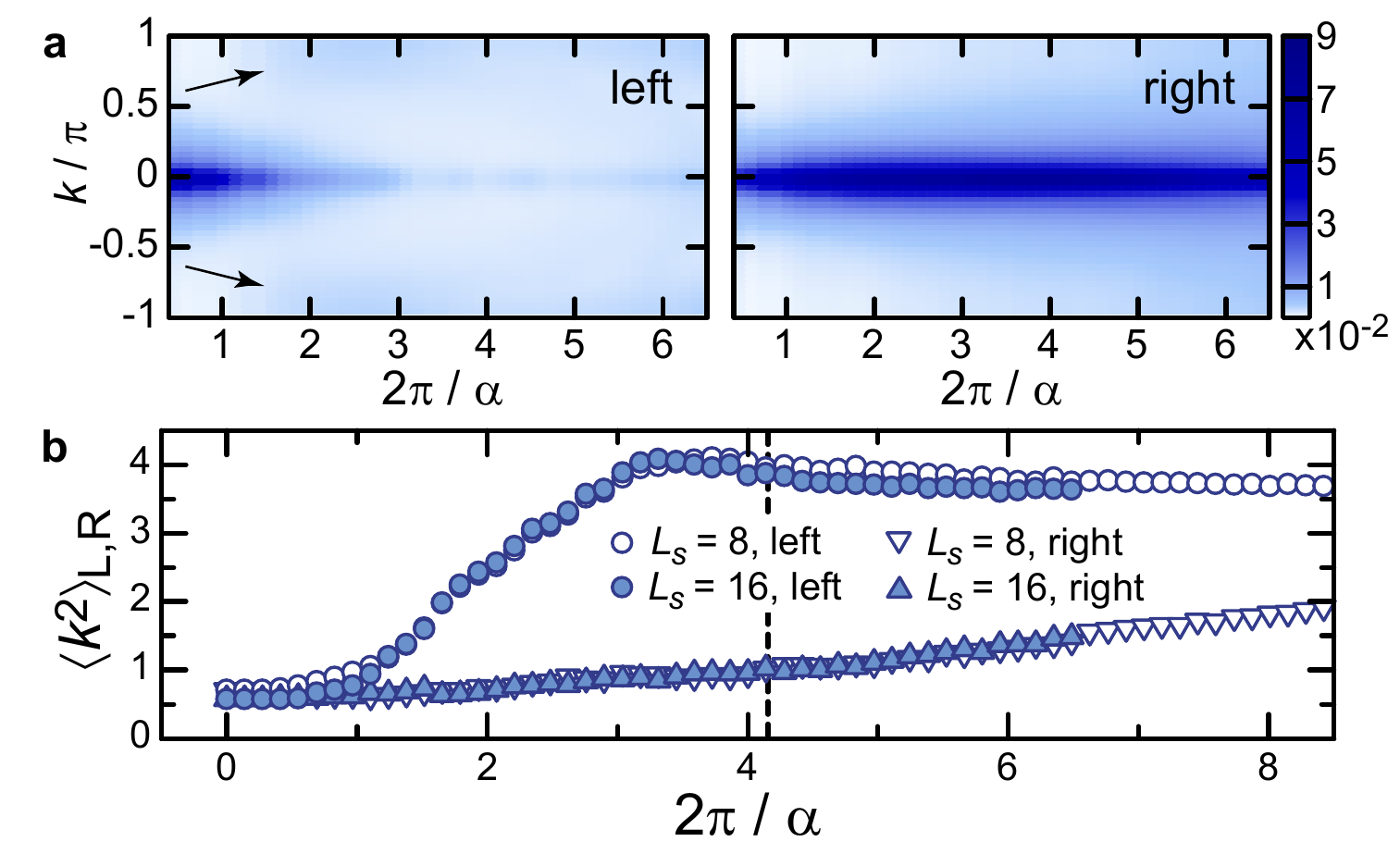}
\caption{\label{fig:MomentumDistrib}
(Color online) 
Broadening of the left- and right-leg momentum distributions during the inverse sweep: (a) Quasi-momentum distributions $n_{k\sigma}$ of the left and the right leg as a function of the reduced sweep time $2\pi/\alpha$ as obtained from DMRG for $n = 0.75$ and $L_s = 16$. The arrows mark peaks in the momentum distributions at $k=\pm \pi$ which appear for $2\pi/\alpha > 1$. (b) The corresponding quasi-momentum width $\langle k^2\rangle_\sigma$ of both legs (filled symbols) plotted together with the results for $L_s = 8$ (open symbols) where slower sweeps are accessible. The dashed vertical line marks the sweep rate for which $n_{\rm R}$ reaches its maximum [see Fig.~\ref{fig:inverseSweep}(c)]. The Hamiltonian parameters are the same as for Fig.~\ref{fig:inverseSweep}.}
\end{figure}  

For very fast sweeps, the left- and right-leg distributions are equal up to a constant factor and therefore have equal widths. Since intra-chain processes are absent on this timescale, each individual Bloch state undergoes a separate two-level LZ transition. Therefore, the momentum distribution in the right leg reduces to $n_{k{\rm R}} = p_\textrm{LZ}(\alpha) n_{k{\rm L}}(t=0)$. As the sweep rate decreases, the distributions of both legs show a completely different behavior. In the left leg, the peak around $k=0$ is more and more depleted, whereas new local maxima emerge near $k=\pm \pi$, signaling that the motion of atoms on neighboring sites in the chains becomes correlated. The width $\langle k^2 \rangle_{\rm L}$ of the quasi-momentum distribution grows correspondingly, reaching a maximum value shortly before the transfer efficiency becomes maximal (dashed vertical line). 

For even slower sweeps, the peaks at $k=\pm \pi$ in the left leg spread out and make the quasi-momentum distribution appear more uniform. This, together with a counterflow of low-momentum components from the right to the left leg, causes the width $\langle k^2 \rangle_{\rm L}$ to decrease. The quasi-momentum distribution of the right leg, in contrast, shows the condensate peak at $k=0$ but the peaks at the edge of the Brillouin zone are almost absent. The associated width $\langle k^2 \rangle_{\rm R}$ increases about linearly for the full range of values $2\pi / \alpha$ covered by our simulations. These results provide numerical evidence for the coupling to low-energy momentum modes being the responsible mechanism for the breakdown of adiabaticity in the inverse sweeps.
  
\textit{Quantum quenches.} A more detailed ``stroboscopic'' understanding of this decay mechanism can be obtained from the quantum quenches denoted by the second case in Eq.~(\ref{deltat}). Since the Hamiltonian is time-independent for $t > 0$, the energy $E$ of the system is a constant of motion throughout the subsequent time-evolution. In Fig.~\ref{fig:quenchMomentum}(a) and (b), we plot the instantaneous values of the fraction $n_{\rm R}(t)$ of particles and the width of the right-leg quasi-momentum distribution $\langle k^2\rangle_{\rm R}(t)$, respectively, for three different values of the bias $\Delta_f$ where the system is again initialized with $\Delta_i$ large and positive. Both observables show oscillations which dampen out towards a non-zero value which depends of $\Delta_f$.

\begin{figure}[tb]
\centering
\includegraphics[scale=1]{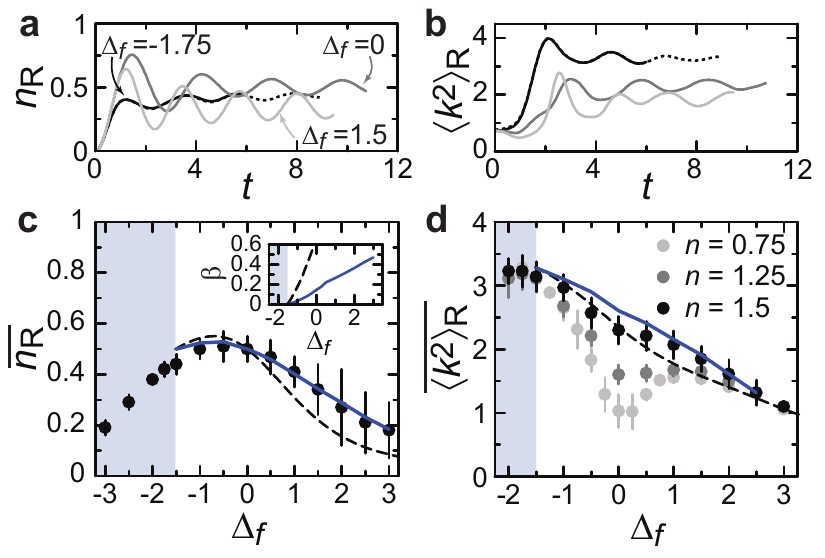}
\caption{\label{fig:quenchMomentum} (Color online) Quantum dynamics after a sudden quench: (a) Time-traces of the right-leg density $n_{\rm R}(t)$ and (b) the width  
$\langle k^2 \rangle_{\rm R}$ of the momentum distribution in the right leg for $n=1.5$ and different values of $\Delta_f$. (c) Long-time value $\overline{n_{\rm R}}$ of the right-leg density versus $\Delta_f$ ($n=1.5$). Points and error bars refer to the time-averaged value and amplitude of the last oscillation, respectively [see Appendix D]. The solid line denotes the right-leg population as obtained for an interacting thermal Bose gas with the same total energy as the initial state in the quenches. The dashed line is the prediction for an ideal Bose gas. The shaded area indicates $\Delta_f \leq -4J_\parallel$. The inset shows the inverse temperatures used in the finite-temperature calculations. (d) Long-time value of the width of the momentum distribution versus detuning for various densities (circles). The solid and dashed lines are the results of the finite temperature calculations as in (b). For all plots, the size of the system is $L_s=16$ and the Hamiltonian parameters are as in Fig.~\ref{fig:inverseSweep}. }
\end{figure} 

In Fig.~\ref{fig:quenchMomentum}(c), we plot the long-time value of the right-leg population $\overline{n_{\rm R}} \equiv n_{\rm R}(t \to \infty)$ as obtained from the time traces for different values of $\Delta_f$. The dots and the error bars represent, respectively, the average value and the amplitude of the last oscillation that we simulate numerically [see Appendix D]. As the bias $\Delta_f$ crosses zero and becomes negative, $\overline{n_{\rm R}}$ reaches a maximum value slightly above 0.5 and then decreases. Any loss of potential energy due to the transfer of atoms to the empty chain must be counterbalanced by an equal increase in kinetic energy. Since this condition can hardly be satisfied when the detuning is large and negative, more and more atoms remain self-trapped in the left well (shaded region). 

For comparison, we performed static DMRG simulations for an interacting Bose gas on the two-leg ladder with bias $\Delta_f$ at thermal equilibrium. We chose the temperature in these simulations such that the total energy matches with the conserved energy $E$ in the quench. The respective results for $n_{\rm R}$ [solid line in Fig.~\ref{fig:quenchMomentum}(c)] are in very good agreement with the long-time values of the dynamical evolution. On the other hand, the same calculation for an ideal Bose gas with $E=-2 J_\parallel n L_s$ (dashed line) shows significant discrepancies, indicating the crucial importance of interactions. The inverse (effective) temperatures $\beta$ found from the matching condition are plotted as inset in Fig.~\ref{fig:quenchMomentum}(c). At the point $\Delta_f=-4 J_\parallel$, $\beta$ becomes zero irrespective of the density or the interaction strength. This implies that quantum quenches with $\Delta_f <-4 J_\parallel$ cannot be associated to an equilibrium thermal state with a finite positive temperature.

In Fig.~\ref{fig:quenchMomentum}(d) we plot the long-time value of the width $\overline{\langle k^2 \rangle_{\rm R}}$ of the right-leg quasi-momentum distribution as a function of $\Delta_f$ and for different densities. Within our numerical accuracy, the quasi-momentum widths in the left and in the right legs coincide for long times, unless $\Delta_f <-4 J_\parallel$, where the dynamics are found to be extremely slow. We again compare the long-time values with the results of the finite-temperature DMRG calculation and find good agreement for the largest density ($n=1.5$), especially away from $\Delta_f=0$. For lower densities, and closer to $\Delta_f = 0$, we observe a less pronounced broadening of the momentum distribution. Here, collisions are much less effective and a complete equilibration cannot be reached before finite-size effects become important [see Appendix D]. At $\Delta_f=-4 J_\parallel$, where the effective temperature diverges, the momentum distribution is flat, corresponding to $\langle k^2 \rangle= \pi^2/3$ for any finite density.

Together with the equilibration of right-leg density, our findings suggest that the system relaxes towards an equilibrium state close to the Gibbs-ensemble associated with the initial state's particle number and total energy. The responsible decay mechanism is the same that leads to the breakdown of adiabaticity and the accompanying broadening of the momentum distribution for slow inverse LZ sweeps.

In conclusion, we have studied numerically both LZ sweeps and quenches in a two-leg ladder system of strongly interacting bosons. For inverse sweeps, we have recovered the breakdown of adiabatic transfer of particles to the initially unoccupied right leg as it was recently found in experiments \cite{Chen:2010}. We have shown that this phenomenon is preceded and accompanied by a fast broadening of the momentum distribution of the initially filled left leg, providing numerical evidence for the coupling to an inner bath of low energy momentum excitations. Finally, we have investigated the underlying decay mechanism by studying quantum quenches of the energy bias. We have found strong evidence for the system to approach a thermal state whose temperature is set by the bias. Our findings provide detailed insight into the dynamics of strongly correlated quantum many-body systems in low dimensions which can be probed in current experiments with ultracold atoms.

This work was funded by the DFG (FOR 635, FOR 801) and the EU (NAMEQUAM).

%


\setcounter{figure}{0}
\setcounter{equation}{0}

\renewcommand\thefigure{S\arabic{figure}}
\renewcommand\theequation{S\arabic{equation}}

\maketitle
\section*{Appendix A -- Time-dependent DMRG simulations}
From our simulations we seek to obtain the final transfer efficiency $n_{\rm R}$ in the limit of $|\Delta_0|,T \to \infty$ in order to compare or findings to the original LZ problem \cite{Landau:1932,*Zener:1932}.
However, for finite values of $|\Delta_0|$ and $T$, when fixing the bias to $-\Delta_0$ for $t>T$, the right-leg population is not constant, but oscillates around an average value. This fact complicates the extraction of a final value for the transfer efficiency especially for fast sweeps and is caused by two reasons. On the one hand, the initial state calculated at $\Delta(t=0) = 100$ with all particles located on the left leg is projected onto the eigenstates of the Hamiltonian~(1) with $\Delta = \Delta_0$ ($|\Delta_0| \ll \Delta(t=0)$). In other words, the initialization of the LZ sweep is already practically non-adiabatic. On the other hand, for finite sweep rates, the sweeps are not fully adiabatic. The final state hence is a partial projection of the initial state onto the eigenstates of (1) with $\Delta = -\Delta_0$. Since these eigenstates consist of contributions from both legs for values of $|\Delta_0|$ not much larger than $1$, coherent population oscillations will occur for $t > T$.

We therefore rescale both the energy  bias $\Delta_0^{\prime} = r \Delta_0$ and the sweep time $T^{\prime} = r T$ by the same factor $r$, which leaves $\alpha=2\Delta_0/T$ unchanged but makes the simulation more adiabatic. In practice in our numerics we choose this 'multiplication' factor between $r =7$ for fast sweeps and $r =2$ for slow sweeps. For each final value of $n_{\rm R}$ in Fig~1(c), we take the average over the, now, much smaller residual oscillations at the end of every single sweep.

\section*{Appendix B -- Choice of Hamiltonian parameters}
In the experiments of Ref.~\cite{Chen:2010}, a superlattice of the form $V_x(x)=V_{xs} \sin^2(2 \pi x/\lambda_{xs})+ V_{xl}\sin^2(2\pi x/\lambda_{xs})$ was created along the $x$-direction by superimposing two standing-wave laser fields with wavelengths $\lambda_{xs} = 765\,{\rm nm}$ and $\lambda_{xl} = 1530\,{\rm nm}$. Additional transverse lattices with wavelengths $\lambda_{y,z} = 844\,{\rm nm}$ and respective depths $V_{y,z}$ completed the setup. For all simulations presented in this Letter, we fix the lattice depths to be $V_{xs} = 15\,E_{\rm rec}^{xs}$, $V_{xl} = 40\,E_{\rm rec}^{xl}$, $V_{y} = 30\,E_{\rm rec}^{y}$ and $V_{z} = 4\,E_{\rm rec}^{z}$ in accordance with a specific measurement in Ref.~\cite{Chen:2010} and calculate the corresponding Hamiltonian parameters from the single particle bandstructure. All lattice depths are given in units of their respective recoil energies $E_{\rm rec}^{i} = h^2/(2 m \lambda_{i})$. 

\section*{Appendix C -- LZ sweeps with few particles in a double well}
As long as $\alpha/2\pi > J_{||}$ in the LZ sweeps, the dynamics of distant rungs in the ladder is still uncorrelated. In this limit, it is helpful to consider the transfer efficiency for a LZ sweep within an isolated double well occupied by $n$ interacting particles, formally corresponding to a ``ladder'' with $L_s = 1$. 

\textit{Ground-state sweeps}. For a strong repulsive on-site interaction $U \gg n$, the ground state undergoes $N$ independent single-particle tunnel resonances, as the bias $\Delta$ is changed \cite{Cheinet:2008,Venumadhav:2010}. At each of these resonances, two Fock-states $\vert n-j,j\rangle$ and $\vert n-j-1, j+1\rangle$ ($j = 0\ldots n-1$) are coupled and the corresponding coupling matrix element for the bosons is $-J_{n,j} \equiv -\sqrt{(n-j)(j+1)} \leq -1$. The propbability to diabatically cross one of these resonances is given by $p_{n,j}(\alpha) = \exp(-2\pi (n-j)(j+1)/\alpha)$. If the $j$-th resonance is crossed diabatically, the double-dot remains in the state $\vert n-j,j\rangle$. When calculating the transfer efficiency of a sweep, we can neglect any higher order crossings between the excited states possibly reached, since the associated coupling matrix elements are typically suppressed as $U^{-\nu}$, where $\nu$ is the order of the respective resonance. Thus, the final transfer efficiency eventually becomes
\begin{equation}\label{eq:anaGSSweep}
  n_{\rm R}(\alpha) = \prod_{j=0}^{n-1} \left[1-\frac{n-j}{n}\,p_{n,j}(\alpha)\right]\,.
\end{equation}

\begin{figure}[tb!]
\centering
\includegraphics[width=0.45\textwidth]{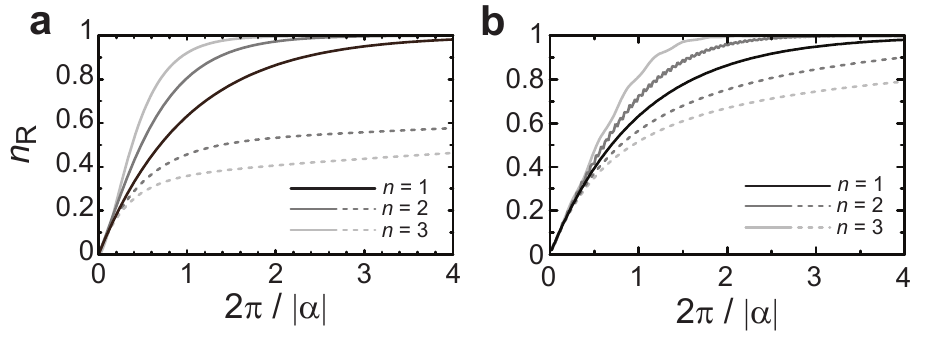}
\caption{\label{fig:DoubleWell}
(Color online) 
Transfer efficiency for a ground-state sweep in a double-dot. (a) Plot of the analytical result for $n_{\rm R}$ in the limit of $U \gg 10$ for $n = 1,2,3$ bosons in the double well in the case of a ground-state sweep (solid lines, Eq.~(\ref{eq:anaGSSweep})) and an inverse sweep (dashed lines, Eq.~(\ref{eq:anaIn vSweep})). (b) Plot of the transfer efficiency as obtained from direct numerical integration of the Schr\"odinger equation for $U=1$ and the same conditions as in (a).}
\end{figure}  

\textit{Inverse sweeps}. Other than the ground state in the double well, the highest excited state undergoes only a single tunneling resonance $\vert n,0\rangle \to \vert 0,n\rangle$ at $\Delta = 0$, where all particles have to change sides simultaneously. In the limit of strong repulsion, this tunnel process is suppressed as $U^{1-n} \ll 1$ ($n > 1$). Therefore, much slower sweeps than in the ground state are needed to achieve full transfer. If this resonance is crossed diabatically and the system remains in state $|n,0\rangle$, it will arrive at another tunnel resonance to $|1,n-1\rangle$ which is of $(n-1)$-th order, as $\Delta$ is increased. Further along this path, the system will face a total of $n$ tunnel resonances of decreasing order and thus for $U \gg 1$ increasing coupling. The coupling matrix element for the $\nu$-th order resonance $|n,0\rangle \to |n-\nu,\nu\rangle$ in leading order is
\begin{equation}
  -J_n^{(\nu)} = \frac{1}{U^{\nu-1}}\,\frac{\nu}{(\nu-1)!}\sqrt{n \choose \nu}
\end{equation}
The probability to follow the eigenstate adiabatically at resonance $\nu$ is thus $1 - p_{n}^{(\nu)}(\alpha) \equiv 1-\exp[-2\pi {J_n^{(\nu)}}^2/\alpha]$. Since this adiabatic crossing results in a final right-well population $\nu$, the overall transfer efficiency becomes

\begin{widetext}
\begin{eqnarray}\label{eq:anaIn vSweep}
  n_{\rm R}(\alpha) &=& (1-p_n^{(n)}(\alpha))  + \frac{n-1}{n}\,p_n^{(n)}(\alpha)(1-p_n^{(n-1)}(\alpha)) + \cdots 
  + \frac{1}{n}\,p_n^{(n)}(\alpha) \cdots p_n^{(2)}(\alpha)(1-p_n^{(1)}(\alpha)) \nonumber\\
  &=& 1 - \frac{1}{n}\,\sum_{\mu=0}^{n-1} \prod_{\nu=n-\mu}^{n} p_n^{(\nu)}(\alpha)\,.
\end{eqnarray}
\end{widetext}

In Fig.~\ref{fig:DoubleWell}(a), we plot the transfer efficiency for both a ground-state sweep and an inverse sweep with $n = 1,2,3$ particles and $U = 10$. In the case of the ground-state sweeps, the enhancement due to bosonic statistics at each resonance leads to an increase of the transfer efficiency with $n$. For the inverse sweeps, however, the transfer efficiency is reduced with increasing particle number due to the higher order effective coupling matrix element. Only the single-particle transfer $|n,0\rangle \to |n-1,1\rangle$ is of first order, so that each of the curves nearly follows the single-particle result until a value of $1/n$, from where on the transfer efficiency is significantly reduced. This qualitative behavior survives for much weaker interactions from a direct numerical integration of the time-dependent Schr\"odinger equation [see Fig.~\ref{fig:DoubleWell}(b)].

\section*{Appendix D -- Long-time values from the quenches and finite-size effects}
To extract the long-time values from the time-traces for $n_{\rm R}(t)$ and $\langle k^2 \rangle$ in the quench scenario [Fig.~3(c) and (d)], we average the traces over the last oscillation period accessible in our $t$-DMRG simulations. For a density of $n = 1.5$, these residual oscillations have only small amplitude and no significant drift is observed anymore [see Fig.~3(a) and (b)]. For low densities, i.~e. $n=0.75$, we are able to follow the dynamics up to times $t \geq T^\ast \equiv L_s/(2J_{||}) \simeq 21$ ($L_s  =16$, $J_{||} = 0.38$), where finite-size effects caused by reflections at the boundaries of the system become important. Here, we estimate the ``asymptotic value'' by taking the average over the oscillation period located around $t = 15 < T^\ast$. Since collisions are rare for such low density, the oscillations amplitudes are slowly damped and the system cannot relax fully within the time $T^\ast$, leading to the pronounced deviations from the results for the thermal state shown in Fig.~3(d).

\end{document}